Article type: Full paper

**Suppressing twin formation in Bi$_2$Se$_3$ thin films**

*N. V. Tarakina*, S. Schreyeck, M. Luysberg, S. Grauer, C. Schumacher, G. Karczewski, K. Brunner, C. Gould, H. Buhmann, R. E. Dunin-Borkowski, and L. W. Molenkamp*

Dr. N. V. Tarakina, S. Schreyeck, S. Grauer, Dr. C. Schumacher, Prof. G. Karczewski, Prof. K. Brunner, Dr. C. Gould, Prof. H. Buhmann, Prof. L. W. Molenkamp
Experimentelle Physik III, Physikalisches Institut and Wilhelm Conrad Röntgen-Research Centre for Complex Material Systems, Universität Würzburg, Am Hubland, D-97074 Würzburg, Germany
E-mail: nadezda.tarakina@physik.uni-wuerzburg.de
Dr. M. Luysberg, Prof. R. E. Dunin-Borkowski
Ernst Ruska-Centre for Microscopy and Spectroscopy with Electrons and Peter Grünberg Institute 5, Forschungszentrum Jülich, D-52425 Jülich, Germany
Prof. G. Karczewski
Institute of Physics, Polish Academy of Sciences, Al. Lotników 32/46, 02-668 Warsaw, Poland

The microstructure of Bi$_2$Se$_3$ topological-insulator thin films grown by molecular beam epitaxy on InP(111)A and InP(111)B substrates that have different surface roughnesses has been studied in detail using X-ray diffraction, X-ray reflectivity, atomic force microscopy and probe-corrected scanning transmission electron microscopy. The use of a rough Fe-doped InP(111)B substrate results in complete suppression of twin formation in the Bi$_2$Se$_3$ thin films and a perfect interface between the films and their substrates. The only type of structural defects that persist in the "twin-free" films is an antiphase domain boundary, which is associated with variations in substrate height. It is also shown that the substrate surface termination determines which family of twin domains dominates.

**1. Introduction**

Bismuth selenide (Bi$_2$Se$_3$) is well-known as a thermoelectric material,[1-3] while also attracting extensive attention as a promising 3D topological insulator (TI). Representing a new state of quantum matter, TIs are appealing for future applications in spintronics and quantum computing, as well as for exploring Majorana fermions.[4, 5] It was first predicted[6] and then shown using angle-resolved photoemission spectroscopy[7] that Bi$_2$Se$_3$ has surface states that reside in the bulk insulating gap (about 0.3 eV) and are protected by time-reversal symmetry. However, in practice

Bi$_2$Se$_3$ often shows high bulk conductivity, which screens the surface-state contribution to transport properties and is an indication that the Fermi level is shifted to the conduction band.[8] Various efforts have been made to tune the Fermi level back into the band gap (e.g., doping of layers[9,10] and the application of top- and back-gating[8]). However, a more effective way to suppress bulk conductivity is to reduce the number of defects in thin films and nanostructures.[11,12] In this work, we focus on the growth of Bi$_2$Se$_3$ thin films using molecular beam epitaxy (MBE). Reports on the growth of Bi$_2$Se$_3$ on different substrates (Si,[13-16] GaAs,[17] Al$_2$O$_3$,[18] InP(111),[19-21] InP(001),[22] and SrTiO$_3$[23, 24]) indicate that mosaicity-tilt, mosaicity-twist and the formation of antiphase domain boundaries (APBs) and twins are the main structural imperfections present in Bi$_2$Se$_3$ films. The use of lattice-matched substrates has been shown to considerably reduce mosaicity twist.[17, 19] Recently, suppression of twinning in Bi$_2$Se$_3$ was demonstrated by the growth of layers on InP(115)[25] and rough InP(111)A[21] substrates. However, in the former study, giant corrugation of the layers resulted in the persistence of mosaicity-tilt, while in the latter study the mechanism of twin suppression was only briefly commented on.

The goal of the present work is to reveal the origin of the formation of different structural defects in Bi$_2$Se$_3$ thin films. We conduct a detailed comparative study of layers grown on InP(111)A and -B terminated flat and rough substrates using reflection high-energy electron diffraction (RHEED), atomic force microscopy (AFM), X-ray reflectivity (XRR), X-ray diffraction (XRD) and probe-corrected scanning transmission electron microscopy (STEM). This choice of substrate reduces the formation of mosaicity twist sufficiently due to an almost perfect lattice match (0.2%) between InP and Bi$_2$Se$_3$. The use of substrates with different terminations and roughnesses allow the factors that define twin formation to be identified, providing conclusions about how twinning can be controlled and suppressed. We believe that our study is relevant not only for Bi$_2$Se$_3$ growth but that it also provides essential insight for obtaining monocrystalline A$_2$B$_3$ (A = Bi, Sb; B = Se, Te) chalcogenide thin films and for realizing desirable electrical properties within this class of materials.

2. Experimental details

Bi$_2$Se$_3$ layers were grown by molecular beam epitaxy in a Createc system with a base pressure of 10$^{-10}$ mbar. The growth temperature for all of the layers was 300°C. The beam equivalent pressure of elemental Se (99.9999%) was 4×10$^{-6}$ mbar and that of Bi (99.9999%) was 2×10$^{-7}$ mbar. These parameters result in a growth rate of ~1 nm per minute (0.017 Å/s). After growth, each sample was cooled to 140°C in a Se flux to prevent Se outdiffusion. Undoped InP(111)B and Fe-doped InP(111)A and (111)B substrates were used. Fe-doped substrates provide high-electrical resistivity, which is necessary for magnetotransport measurements.

Annealing of the substrates above 300°C, which leads to a flattening of the surface, was performed at the presence of a Se flux to prevent phosphorus outdiffusion. For the annealed substrates the natural oxide layer was removed by thermal desorption. For substrates that had not been annealed above the growth temperature of 300°C, the natural oxide layer was removed by dipping them into 50% hydrofluoric acid. Growth was monitored by RHEED using a CCD camera. After growth, the samples were characterized using a DME DualScope 95–50 atomic force microscope and a Philips X'Pert MRD diffractometer, with which XRD and XRR measurements were performed.

TEM studies were carried out using a probe-corrected FEI Titan 80-300 (S)TEM operated at 300 kV. Cross-sectional TEM specimens were prepared in two steps. First, samples were thinned down to a thickness of ~100 nm by focused ion beam milling using an FEI Helios Nanolab Dual Beam system. We used a 30 kV Ga$^+$ ion beam, currents of between 0.28 nA and 28 pA and a 1° incident angle. Final low-kV cleaning was performed using a Fischione NanoMill system at 900 eV and 160 μA. Simulations of STEM images made using the JEMS software package.[26]

## 3. Results and discussion

### 3.1. Bi$_2$Se$_3$ grown on flat InP(111)B

Bi$_2$Se$_3$ layers grown on flat InP(111)B were found to suffer from two structural imperfections: poor crystal quality of the interface and the presence of twin domains. This observation agrees well with previous studies.[19, 20] However, the use of a probe-corrected (S)TEM allows important details about these defects to be revealed.

Cross-sectional high-angle annular dark-field scanning-transmission electron microscopy (HAADF-STEM) images show that the 'poor-crystalline quality' interface layer in fact consists of small crystalline domains of $Bi_2Se_3$, which are not always perfectly aligned to the substrate (**Figure 1**). Small misalignments cause a lowering of the contrast and the apparent resolution of the HAADF-STEM images, making misaligned areas look amorphous.

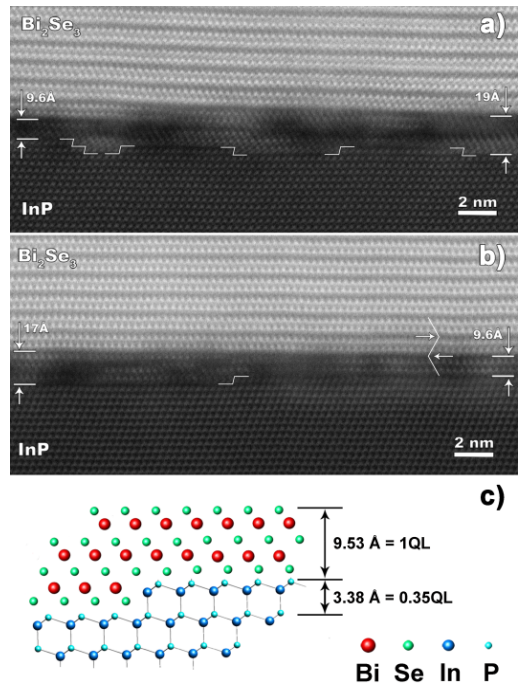

**Figure 1.** (a,b) Cross-sectional HAADF-STEM images of the interface between a $Bi_2Se_3$ film and a flat InP(111)B substrate recorded from two different regions of the film. (c) Schematic representation of a "$Bi_3Se_4$" cluster formed on a diatomic step of InP. Horizontal arrows and the kinks in zig-zag lines indicate positions of boundaries between lamellar twins. Step-shaped lines outline diatomic steps on the surface of InP.

The presence of small domains at the interface suggests that many nucleation points are present at the very first step of growth. The thickness of the poor crystalline interface layer was found to vary between 0 and 2 quintuple layers (QLs). Each of the domains at the interface forms a QL and misalignment mainly occurs at areas where domains meet and try to merge. There are two difficulties in this process. First, the 'flat' InP(111)B substrate (root-mean-square surface roughness ($R_{RMS}$) is 0.1 nm) is not in fact atomically flat, but has diatomic surface steps (DS) of 3.38 Å (0.35

QL) as shown on Figure 1. In order to overcome the presence of diatomic steps on the surface of the substrate, an additional cluster with chemical composition $Bi_3Se_4$ forms (Figure 1).

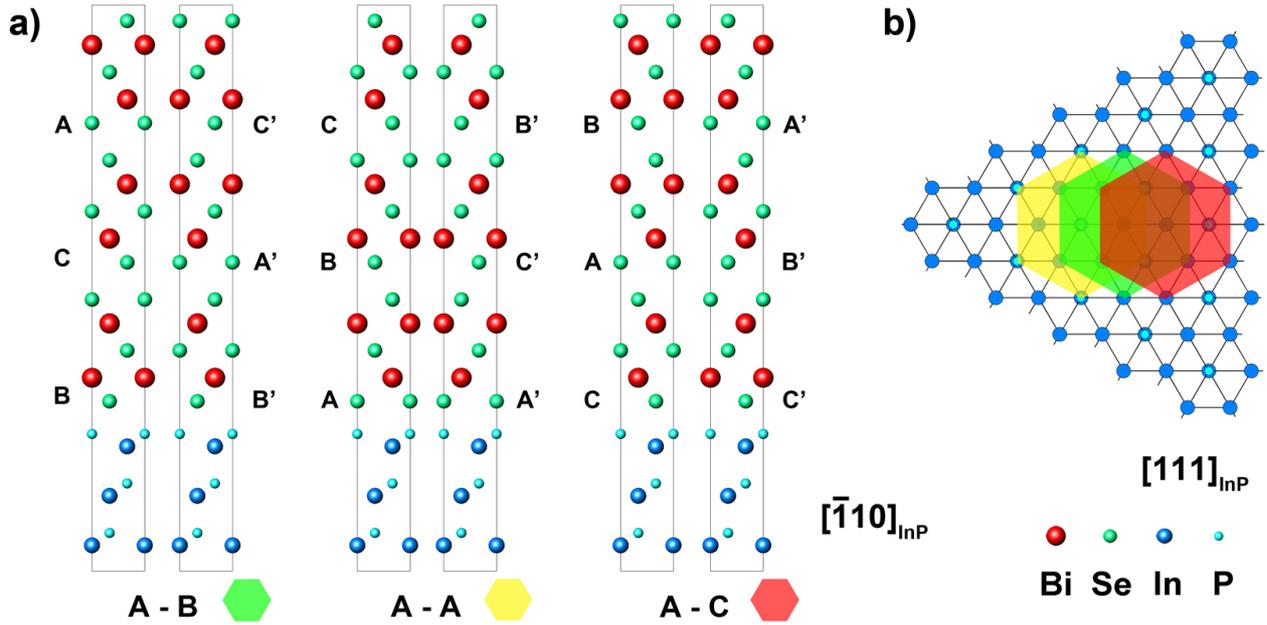

**Figure 2.** Representation of the different stackings that are possible between an InP(111)B substrate and $Bi_2Se_3$ layers and the corresponding twins. Capital letters next to the $Bi_2Se_3$ structure denote the position of the first atomic layer in the QL, while primed capital letters denote the first layer in the QL of the twin. The top phosphorus layer of InP is always set to be in the A position. (a) View along the [-110] direction of InP; (b) view along the [111] direction of InP.

The formation of such seven-layer $Bi_3Se_4$ lamellae has also been found in bulk $Bi_2Se_3$,[27, 28] as well as in both bulk and thin films of $Bi_2Te_3$ (but not at the interface).[27, 29] The height of one and even two diatomic steps is not sufficient to match the height of a domain (1QL = 9.53 Å), which results in the formation of clusters with different chemical compositions, as well as in the presence of misalignment and defects at the interface. Second, even if two nucleation points form on a perfectly flat area, there are still three crystallographically equivalent ways for the $Bi_2Se_3$ layers to grow on an InP(111)B surface. Following the common notation for cubic close packing (*ccp*) atomic sequences and assuming the upper phosphorus layer of the substrate to be an A layer, three different positions for the first Se layer of a QL of $Bi_2Se_3$ are possible: A-B, A-A and A-C (**Figure 2**). Each of these positions results in the formation of a twin, depending on how the second layer of the QL is formed, for example, A-A-B or A-A-C (Figure 2). Thus, even if only one type of stacking sequence between the substrate and the Bi$_2$Se$_3$ is energetically preferable, two twin domains can form with equal probability. We observed such twin domains at the interface. When they meet they form a twin boundary, which introduces defects into the film (**Figure 3**).

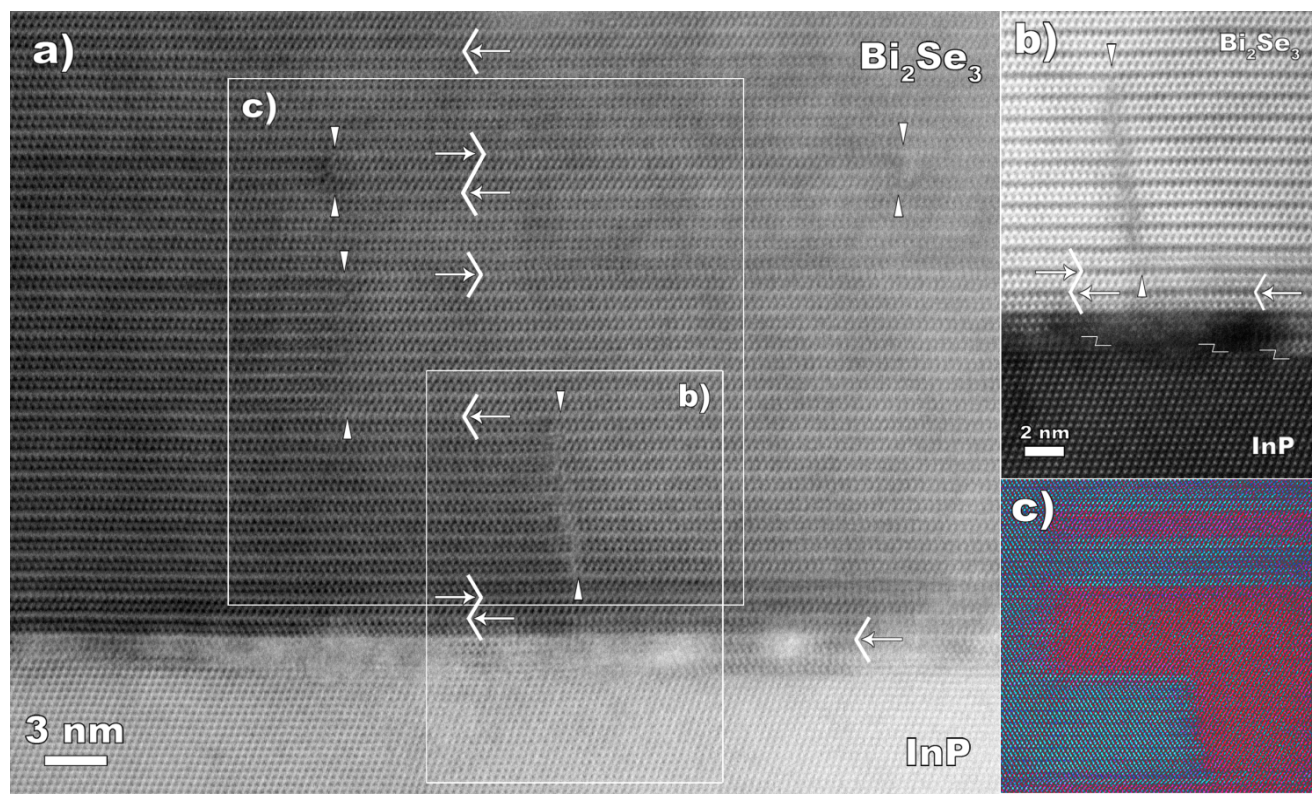

**Figure 3.** Cross-sectional images of an area of a Bi$_2$Se$_3$ film grown on a flat InP(111)B substrate: (a) overview bright-field (BF) STEM image; (b) HAADF-STEM image showing an enlargement of image (a) at the interface region; (c) pseudo-dark-field image of the central part of image (a) with artificial colours used to display two families of twin domains. Small vertical arrows mark positions of twin boundaries, formed perpendicular to the substrate, horizontal arrows and kinks in zig-zag lines mark positions of twin boundaries, formed parallel to the substrate. Step-shaped lines outline diatomic steps on the surface of InP.

For both reasons (the formation of domains with height of one QL at the beginning of growth and twin formation) we conclude that '2D information' passed on to the film by a flat substrate is not sufficient for realizing the controlled growth of Bi$_2$Se$_3$. The question that arises is whether growth can be controlled by employing rough 3D-structured substrate surfaces. To check this possibility, we studied rough InP(111)B Fe-doped substrates.

## 3.2. Bi$_2$Se$_3$ grown on rough Fe-doped InP(111)B

In order to study the influence of roughness on the quality of the Bi$_2$Se$_3$ films, a series of layers was grown on rough Fe-doped InP(111)B (heated to a growth temperature of 300°C) and on the same substrates following annealing prior to growth at 570, 620, and 730 °C. Annealing at high temperatures leads to a flattening of the substrate, which can be observed directly using RHEED (**Figure 4** (a)).

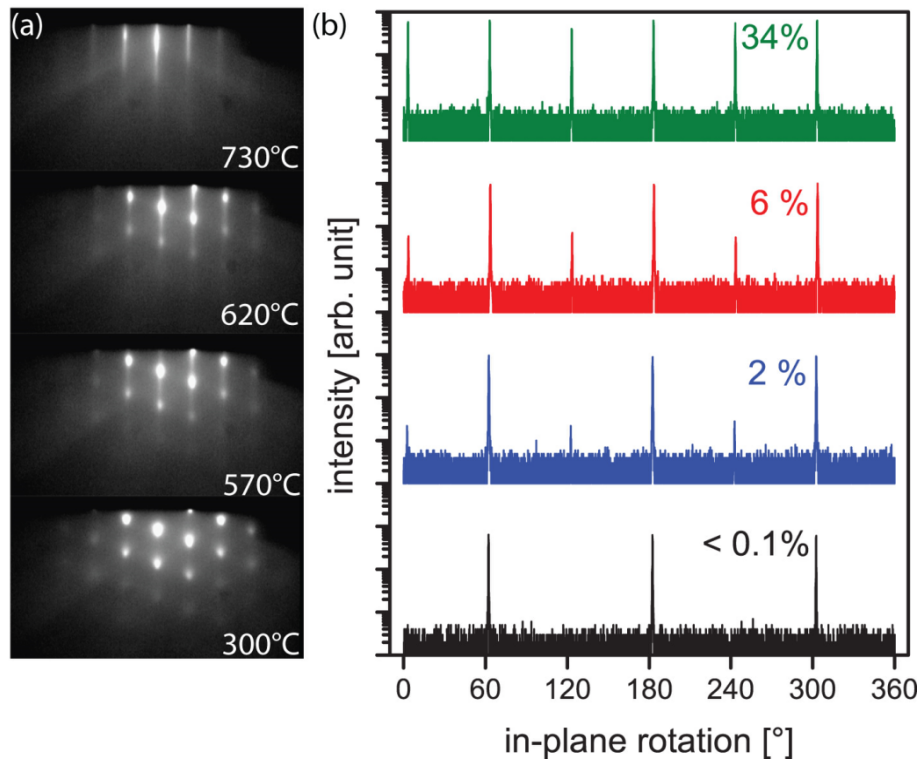

**Figure 4.** (a) RHEED patterns recorded from the InP substrate for different annealing temperatures. (b) Pole scans of the {015} reflections of ~100 nm Bi$_2$Se$_3$ for different annealing temperatures of the InP(111)B substrates; the indicated twin volume of the crystal is determined by the peak area of the twin triplet compared to the total measured peak area. The InP {002} reflections (not shown) occur at the same in-plane angles as the non-suppressed peak triplet. A logarithmic scale is used in the plots.

For a substrate temperature of 300 °C a spotty RHEED pattern is visible, due to the presence of a rough surface resulting in 3D diffraction conditions for the electrons. For higher temperatures, both spots and lines appear on the RHEED pattern. Gradual heating of the substrate leads to a decrease in the intensity of the spots and a sharpening of more intense line features. At 730 °C, the spots have vanished completely, indicating the formation of a smooth surface (2D diffraction conditions).

XRD pole scans of the {015} reflections of the resulting series of $Bi_2Se_3$ layers are shown in Figure 4 (b). For the substrates that had been annealed at high temperatures, six {015} reflections are visible. The intensity of one triplet of peaks decreases with decreasing substrate annealing temperature. For the rough substrates that had only been heated to 300 °C without subsequent annealing, only one triplet remains (a possible second triplet has an intensity of below 0.1%), indicating complete suppression of one family of twin domains.

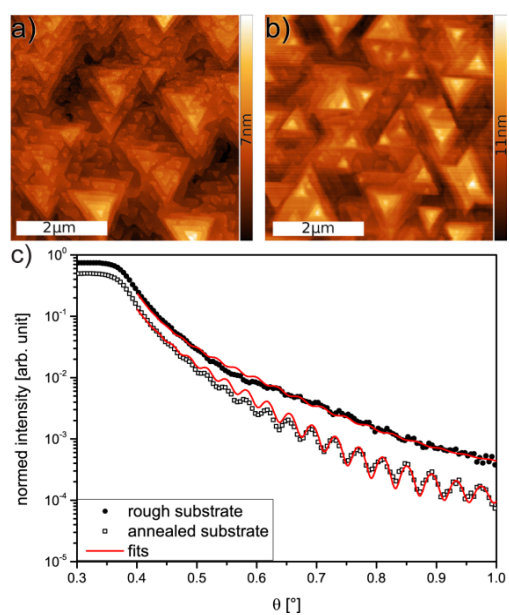

**Figure. 5** AFM images of the $Bi_2Se_3$ surface for (a) a rough substrate and (b) a substrate that had been annealed (at 730 °C). The z scale corresponds to the maximum height. In (c), XRR measurements recorded from these samples are shown, together with the fits (red) that were used to obtain the interface roughness.

In order to determine the interface roughness, AFM and X-ray reflection measurements were performed. AFM images of the surfaces of the $Bi_2Se_3$ films that had been grown on a rough as-bought substrate and a substrate annealed at 730°C are shown in **Figure 5** (a) and (b), respectively. On the annealed substrates, the Bi$_2$Se$_3$ layers display two types of pyramidal domains, which are rotated by 60° with respect to each other. In contrast, the layers on the non-annealed (rough) wafers exhibit a uniform orientation of their domains, implying the absence of twins, which is consistent with the data obtained from XRD measurements.

AFM measured values of the R$_{RMS}$ were found to be 1.1 and 1.6 nm for Bi$_2$Se$_3$ grown on as-bought rough substrates and substrates that had been annealed (at 730°C) substrates, respectively. These values were used to fit the XRR curves (Figure 5(c)). Layers that had been grown on the annealed substrate display distinct oscillations in the XRR curves as compared to layers grown on the rough substrate. The values of the root-mean-square roughness of the InP/Bi$_2$Se$_3$ interfaces (deduced from the fit) were 2.1 and 0.3 nm for the as-bought and annealed substrates, respectively. The XRD and AFM measurements show that the use of a rough substrate enables complete suppression of twin formation. In order to check the interface quality and to obtain details about the mechanism of twin suppression, high-resolution STEM (HRSTEM) studies were performed.

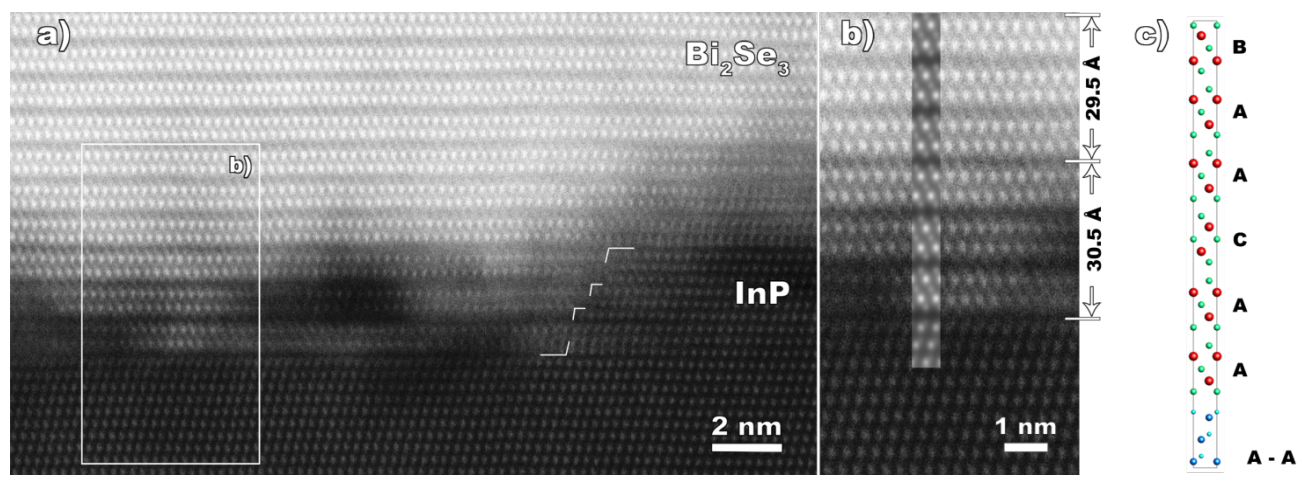

**Figure 6.** Cross-sectional HAADF-STEM images of an area of a Bi$_2$Se$_3$ film grown on a rough InP(111)B substrate: (a) overview image showing a perfect interface and the absence of twins; (b) enlargement of image (a) at the interface region, with a simulated HAADF-STEM image inserted. A difference in contrast between the experimental HAADF-STEM image and the simulated image is present since roughness has been not included in the simulation; (c) structural model of image (b)

with Bi, Se, In, P shown as red, green, light-blue and blue spheres, respectively. Step-shaped lines outline steps on the side surface of InP hollows.

HRSTEM images of the $Bi_2Se_3$ film grown on the rough substrate reveal a high quality interface, the absence of twin boundaries and the absence of microdomains (**Figure 6**). The observed slight fading of the contrast in the images at the interface is caused by overlap of the substrate and the film at the same regions in the cross-sectional images due to roughness of the substrate. The depth of the hollows observed on STEM images varies from 5DS to 15DS. Since the sides of the hollows are higher than the height of a QL (9.6 Å), they behave as additional {1-11} surfaces, making an angle of 70.5 degrees with the substrate surface. These additional surfaces have the same crystal symmetry as the (111) surface of the substrate but opposite terminations. It has been shown in several previous publications that $Bi_2Se_3$ tends to nucleate at the ascending steps of the substrates.[21, 25] If it can be assumed that this behaviour also holds for rough surfaces, bearing in mind the symmetry of the substrate surface and of the side surfaces of the hollows, one can conclude that two of these surfaces (the substrate surface and the side surface of the hollow) define the alignment of the QL layers and the stacking within a QL, providing the '3D information' that results in unique layer stacking. Additional proof of the fact that the $Bi_2Se_3$ layers accommodate the side surfaces of the hollows is the "elongation" of the $Bi_2Se_3$ unit cell within the cavity (Figure 6). Thus, in Figure 6, the first unit cell (the 3 first QLs) is in the cavity and has a *c* lattice parameter of 30.5 Å, which is almost equal to the height of twelve diatomic steps of InP (12 DS = 30.42 Å). The next unit cell in the image is less stressed due to the presence of side surfaces and has a *c* unit cell parameter of 29.5 Å. The third unit cell has *c* = 28.7 Å, which is almost equal to the lattice constant found for $Bi_2Se_3$ films grown on a flat surface[14, 19] and in bulk form.[27, 30] The formation of a bond that is stronger than the Van der Waals bond between the side surface of the hollow and the QL probably accompanies such accommodation.

During growth of the first QL of $Bi_2Se_3$ on rough substrates, the RHEED patterns initially showed a mixture of spots and line features, while after the deposition of 5 QL the patterns became completely streaky. This observation also suggests that the Bi$_2$Se$_3$ film starts to grow within the hollows of a rough substrate.

We performed simulations of the interface regions observed in several HRSTEM images. We found that the first Bi$_2$Se$_3$ QL stacks onto the surface of the InP rough substrate in an A-A sequence (Figure 6 (b, c); Figure 2), which may be defined by the 3D structure of the substrate. Moreover, none of the simulated regions display the *ccp* stacking of QLs present in the monocrystalline bulk compound,[27, 30] but rather show different polymorphic stacking orders (Figure 6 (c)). It is worth mentioning that inside a QL layer the *ccp* arrangement is preserved. Recently, density-functional theory (DFT) calculations have been used to show that the presence of stacking faults in Bi$_2$Se$_3$ layers, which lead to the appearance of different polymorphs, can influence topological properties.[31]

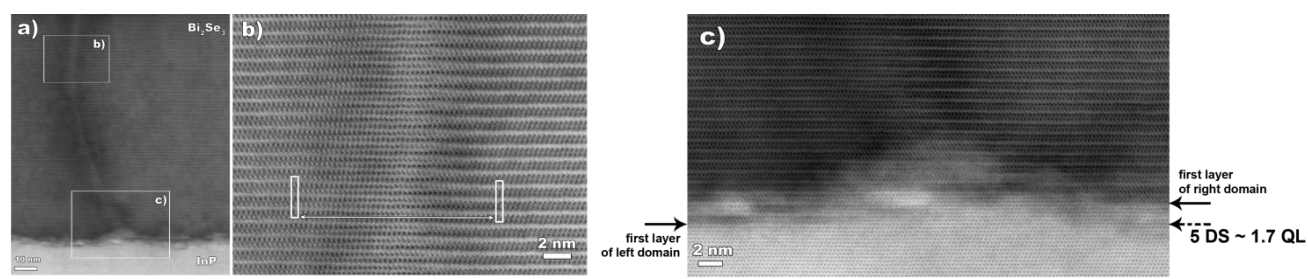

**Figure 7.** Cross-sectional STEM images of an antiphase domain boundary of a Bi2Se3 film grown on a rough InP(111)B substrate: (a) overview BF-STEM image, (b) HAADF-STEM image of the upper part of the antiphase domain boundary, (c) enlargement of image (a) at the interface region.

The only structural defects that were observed in the film are antiphase domain boundaries (translation domains).[32] The presence of this type of defect results from the varying substrate height. The first QLs of different grains start to grow at different levels, so that when they meet they form a grain boundary with a shift of the unit cell of about one third of a QL in the *c*-direction with respect to each other (**Figure 7**).

Such a shift approximately corresponds to the height of one diatomic step. The same type of defect has been observed in Bi$_2$Te$_3$ grown on Si(111)[33] and Bi$_2$Se$_3$ grown on epitaxial graphene/SiC(0001) thin films.[34] In the latter study, the electrical behaviour of the Bi$_2$Se$_3$ films across an antiphase domain boundary was analysed using scanning tunnelling microscopy (STM). The authors showed that such a defect is undesirable and influences the electrical properties of the film negatively. We propose that the use of a substrate with controlled surface roughness (nanopatterning), in particular with hollows of equal depth, could be used to obtain single-crystal films without antiphase domain boundaries.

In spite of the presence of antiphase domain boundaries the samples grown on the rough substrates showed an up to 90% reduction of the carrier density compared to that obtained for $Bi_2Se_3$ layers grown on flat InP(111)B.

### 3.3. $Bi_2Se_3$ grown on rough InP(111)A

In order to clarify whether roughness is the only factor that controls twin suppression, and hence the improvement of interface quality, we performed similar experiments using Fe-doped InP(111)A substrates.

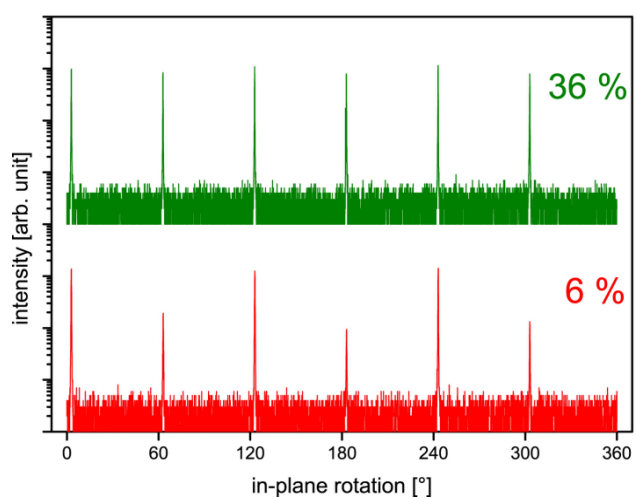

**Figure 8.** Pole scans of the {015} reflections of ~100 nm $Bi_2Se_3$ for growth on rough (red) and flat (green) InP(111)A substrates. The twin volume on A-type substrates (indicated) also demonstrates twin suppression on rough substrates.

AFM, XRR, and XRD measurements showed a similar tendency to that found for B-terminated InP substrates and in particular the suppression of one family of twins with increasing substrate surface roughness (**Figure 8**). It is worth mentioning that the only difference observed is the selection of the triplet of $Bi_2Se_3$ {015} reflections, suppressed by surface roughness. For an A-terminated substrate, the intensity of the triplet, which occurs at the same in-plane rotation angles in the XRD pattern as the InP {002} peaks, decreases with increasing substrate surface roughness, while for the B-type substrate it is the other triplet that is completely suppressed by the roughness. Based on this observation it is possible to conclude that the presence and the number of twins are controlled only by the degree of substrate surface roughness.

## 4. Conclusions

The microstructural properties of $Bi_2Se_3$ topological-insulator thin films are highly dependent on the choice of the substrate. In this work, we have shown that growth using molecular beam epitaxy on rough Fe-doped InP(111) substrates leads to the formation of high quality thin films, with very low mosaicity-twist and with complete suppression of twins in the $Bi_2Se_3$ thin films. No extra layer was observed at the interface between the film and the substrate. The suppression of twins results in a reduction of the carrier density of nearly an order of magnitude, compared to values obtained for twinned $Bi_2Se_3$ layers. We also showed that the substrate surface termination (A or B) defines which family of twin domains dominates. The only type of structural defects that remains in the films are the antiphase domain boundaries associated with the variation in the substrate height.

**Acknowledgements**

This work has been funded by the EU ERC-AG Program (project 3-TOP), by the Alexander von Humboldt Foundation and by The Helmholtz Virtual Institute for Topological Insulators (VITI). N.V.T. acknowledges funding by the Bavarian Ministry of Sciences, Research and the Arts. The authors are grateful to D. Meertens for assistance with TEM sample preparation. N. V. Tarakina and S. Schreyeck contributed equally to this work.